\documentclass[aps,%
groupedaddress]{revtex4-2}
\usepackage{comment}
\usepackage{ulem}
\usepackage{color}
\usepackage{hyperref}
\usepackage{float}
\usepackage{amsmath,amsfonts,amssymb,epsfig,graphicx}
\usepackage{slashed}
\usepackage{lineno}

\newcommand{\MGMCatNLO}{MadGraph5\_aMC@NLO} 

\newcommand{\pt}{\ensuremath{p_\mathrm{T}}}

\newcommand{\PZ}{\ensuremath{\mathrm{Z}}}
\newcommand{\PH}{\ensuremath{\mathrm{H}}}
\newcommand{\PW}{\ensuremath{\mathrm{W}}}

\newcommand{\mup}{\ensuremath{\mu^+}}
\newcommand{\mum}{\ensuremath{\mu^-}}
\newcommand{\elp}{\ensuremath{\mathrm{e}^+}}
\newcommand{\elm}{\ensuremath{\mathrm{e}^-}}
\newcommand{\nue}{\ensuremath{\nu_{\mathrm{e}}}}
\newcommand{\nuae}{\ensuremath{\tilde{\nu}_{\mathrm{e}}}}
\newcommand{\num}{\ensuremath{\nu_{\mu}}}
\newcommand{\nuam}{\ensuremath{\tilde{\nu}_{\mathrm{\mu}}}}
\newcommand{\eV}{\,\text{eV}}

\newcommand{\GeV}{\,\text{GeV}}
\newcommand{\TeV}{\,\text{TeV}}

\newcommand{\fbinv}{\mbox{\ensuremath{~\mathrm{fb^{-1}}}}}
\def\lag{{\cal L}}
\newcommand{\confirm}[1]{{\color{black}#1}}
\newcommand{\mn}{{\rm M}_{\rm{N}}}

\graphicspath{{./pic/}}
\begin{document}

\title{Richness out of smallness: a Possible Staged Blueprint on Future Colliders}


\author{Meng \surname{Lu}$\ast$$^{1}$}
\email[]{meng.lu@cern.ch}

\author{Qiang \surname{Li}$\ast$$^{2}$}
\email[]{qliphy0@pku.edu.cn}

\author{Zhengyun \surname{You}$^{1}$}
\email[]{youzhy5@mail.sysu.edu.cn}

\author{Ce \surname{Zhang}$^{2}$}
\email[]{ce.zhang@pku.edu.cn}

\affiliation{1. School of Physics, Sun Yat-Sen University, Guangzhou 510275, China}
\affiliation{2. State Key Laboratory of Nuclear Physics and Technology, School of Physics, Peking University, Beijing, 100871, China}

\begin{abstract}

Novel collision methods and rich phenomena are crucial to keeping high-energy collision physics more robust and attractive.
In this document, we present a staged blueprint for future high-energy colliders: from neutrino-neutrino collision, neutrino-lepton collision to electron-muon and muon-muon collisions.
Neutrino beam from TeV scale muons is a good candidate to enrich high-energy collision programs and can serve as a practical step toward a high-energy muon collider, which still requires tens of years of R$\&$D.
Neutrinos-neutrinos collision provides a promising way to probe heavy Majorana neutrinos and effective neutrino mass; neutrino and antineutrino annihilation into Z boson has a huge cross-section at 10K pb level; leptons-neutrinos collision benefits W boson mass precision measurements. With only a minimal amount of integrated luminosity, one can envision the ``Richness out of smallness''. 
This document summarizes the current status and the roadmap towards the muon-muon collider with less challenging techniques required through intermediate facilities, where a wide variety of physics goals could be achieved. A (preparatory) laboratory on novel colliders could attract vast international interests and collaborations.

\end{abstract}

\maketitle

\section{Introduction}
\label{introduction}

The discovery of the Higgs boson at the LHC in 2012~\cite{C1} symbolizes that particle physics is entering a key period. On one hand, direct searches for new physics beyond the Standard Model (BSM, such as supersymmetry and extra dimensions) through the Higgs portal receive intense attention. On the other hand, rich progress has been made on heavy flavor and electroweak measurements from the LHC and other experiments, which deepens the scope of precision tests on the SM, and stimulates indirect searches for BSM with the effective field method in a bottom-up approach~\cite{C2}.

Recent years have witnessed several significant anomalies or hints of possible new physics BSM. First, the LHCb Collaboration, in a test of lepton flavor universality using B$\rightarrow$K$ll$, reports a measurement that deviates by 3.1 standard deviations from the SM prediction~\cite{C14}. Second, the latest result from the Muon $g-2$ Experiment at Fermilab has pushed the world average of the muon anomalous magnetic moment measurements to 4.2 standard deviations away from the SM prediction~\cite{C15}. Most recently, the CDF II collaboration has reported a measurement of the W gauge boson mass~\cite{C18}, ${\rm M}_{\rm W}^{\rm CDF} = 80.433 \pm 0.009 \,\, {\rm GeV}$, which is $7.2$ standard deviations away from the SM prediction of ${\rm M}_{\rm W}^{\rm SM} = 80.357 \pm 0.006 \,\, {\rm GeV}$~\cite{C19}. Numerous theoretical studies attempt to accommodate these anomalies, which may or may not require a modification of the SM.

In the next stage, the LHC will enter the HL-LHC phase after 2025-2027~\cite{C3,C4,C5}, and will collect in total around 3000 fb$^{-1}$ of data in a period of 10 years, which can help deepen our understanding of fundamental physics. In addition, HEP communities have had intense discussions on the target and strategy for future colliders (see e.g.~\cite{C6,C7}). 
Various options include, electron-positron collider at the collision energy from 250\GeV~to 3\TeV~\cite{C8,C9,C10,C11,C12}, hadron collider at 100\TeV~scale, and TeV scale muon colliders~\cite{C13}, etc. These future colliders are aiming at precision measurement of Higgs properties and searching for new physics at higher energy scales.
The International Linear Collider (ILC) costs around 10B dollars; the 100 km double-ring circular electron-positron collider (CEPC) and the Future Circular Collider (FCC) cost less but are space-consuming due to energy loss from synchrotron radiation.

Muons suffer less synchrotron energy loss by 8 orders of magnitude than electrons and positrons, which leads to the fact that a TeV scale muon collider can be kept as small as O(km) in circumference. muon collider has a much cleaner environment and larger effective center of mass energy with respect to the collision energy than the hadron collider. It is also sensitive directly to muon-related new physics. Recently, due to the LHCb lepton flavor universality and Fermilab muon $g-2$ anomalies, interest in muon colliders has revived~\cite{C16}. 

It is generally believed that a muon collider still needs decades of research and challenging development, especially, on how to achieve high quality (intensity and emittance) beam and mitigate the beam-induced background (BIB) effects from muon decays.
Muon beams are usually achieved in proton or positron interactions on target, which have both pros and cons on beam cooling or intensity. BIB is crucial in the physics program at a muon collider~\cite{C17}, for which usually a nozzle or a timing detector is introduced to mitigate such effect.

TeV scale muon beams emit bunches of collimated decay products and thus can provide neutrino beams, which have a great potential for application in high-energy collision programs.
With only a small amount of integrated luminosity, interesting phenomena from neutrino collisions can be observed towards the ``Richness out of smallness''.
For example, neutrino and antineutrino annihilate into Z boson with a huge cross-section at 10K pb level. It also opens doors to many new topics, such as searching for resonances that decay into neutrinos. 
One can also collide neutrinos with neutrinos to probe heavy Majorana neutrinos (HMN) and effective neutrino mass. Furthermore, leptons and neutrinos can collide into W bosons, which can be also observed with a very small amount of integrated luminosity and benefit W boson mass precision measurements.  

Through all these stages, a rich variety of physics goals could be achieved, and may also be useful intermediate steps as the study of the challenging technologies towards the muon-muon collider by the energy frontier community.

Section~\ref{physics} elaborates the physics potentials of these collision schemes. Technical considerations are presented in Section~\ref{tc}. Status and prospects are presented in Section~\ref{statusprospects}. The work is summarized in Section~\ref{summary}.

\section{Physics Potentials}
\label{physics}

Neutrinos are among the most abundant and least understood of all particles in the SM that make up our universe. Observation of neutrino oscillations confirms that at least two types of SM neutrinos have a tiny, but strictly nonzero, mass. The upper limits on each neutrino mass come from many experiments including cosmology and direct measurements~\cite{C19}. For example, the Karlsruhe Tritium Neutrino (KATRIN) experiment~\cite{C20} finds an upper limit $m_{\nuae} < 0.8\,\eV$ at the 90\% C.L. for the electron anti-neutrino $\nuae$.  On the other hand, the direct mass limit on electron neutrinos and muon neutrinos are relatively much looser~\cite{C19}. The simplest formalism in which neutrino masses can arise is through a dimension-5 operator as shown by Weinberg~\cite{C21}, which extends the SM Lagrangian with
\begin{align}
\lag_5 = \left( \ {\rm C}_5^{\ell\ell'}/\Lambda \right) \big[\Phi\!\cdot\! \overline{L}^c_{\ell }\big] \big[L_{\ell'}\!\!\cdot\!\Phi\big], 
\label{wbo}
\end{align}
where $\ell ,~\ell'$ are the flavors of the leptons, which can be electrons, muons or taus; $\Lambda$ is the relevant new physics scale; ${\rm C}_5^{\ell\ell'}$ is a flavor-dependent Wilson coefficient; $L_\ell^T=(\nu_\ell,\ell)$ is the left-handed lepton doublet; and $\Phi$ is the SM Higgs doublet with a vacuum expectation value $v=\sqrt{2}\langle\Phi\rangle\approx 246\GeV$. The Weinberg operator generates the Majorana neutrino masses as $\confirm{m_{\ell\ell} = C_5^{\ell\ell} v^2/\Lambda}=\left|\sum_i U_{\mathrm{\ell}i}^2 m_i\right|$, and introduces lepton number violation (LNV).

The ultraviolet (UV) completion of the Weinberg operator can be realised in the context of ``see-saw'' models~\cite{C22}, assuming the existence of hypothetical heavy states, for example the HMN in the type-I seesaw model. Searches for neutrinoless double beta decay in the decays of heavy nuclei have placed strong limit, i.e.,  $m_{\ell\ell}<0.08-0.18\eV$ at the 90\% C.L.~\cite{C23}.

\begin{figure}
    \centering
    \includegraphics[width=0.45\textwidth]{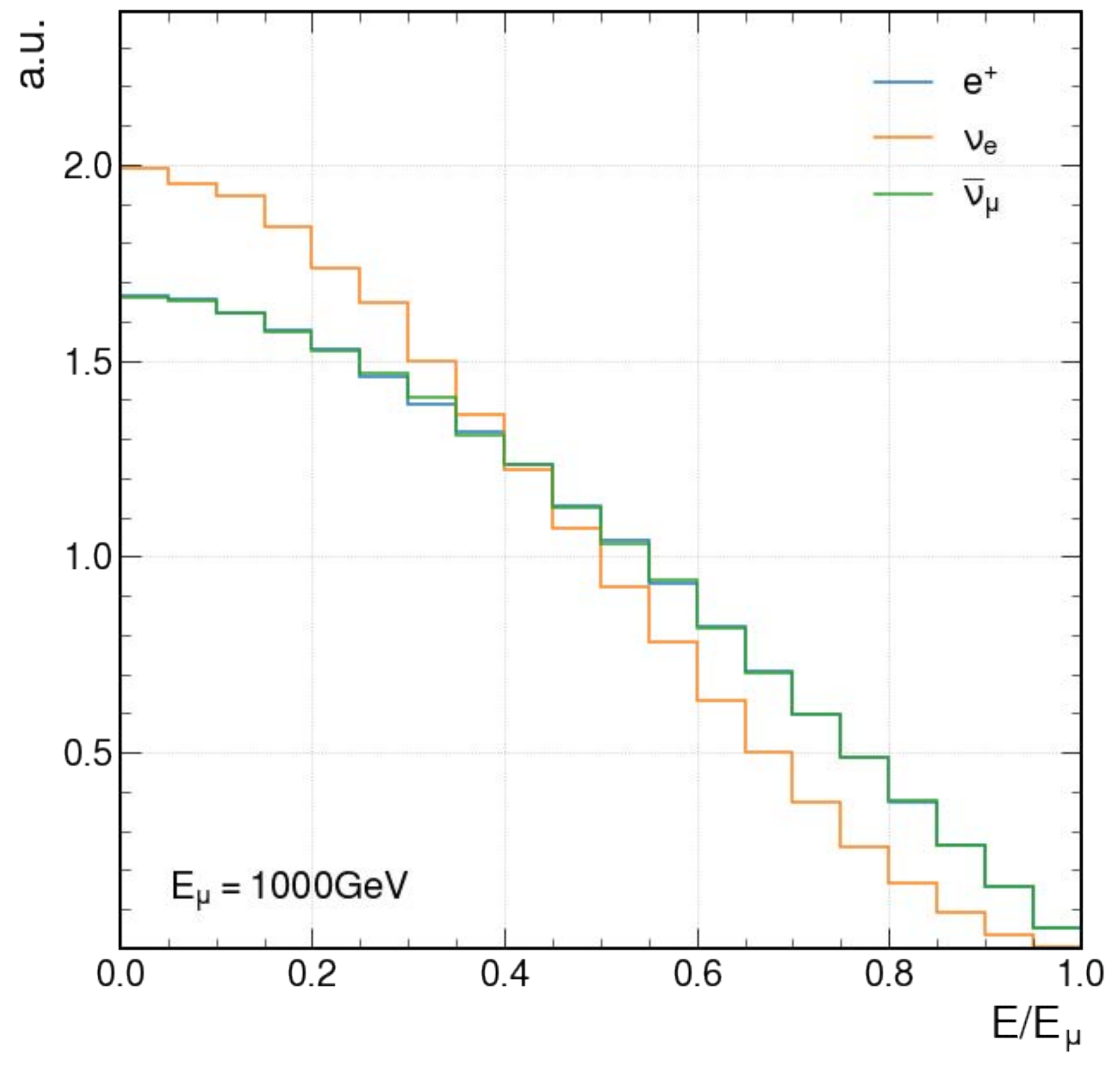}
    \includegraphics[width=0.45\textwidth]{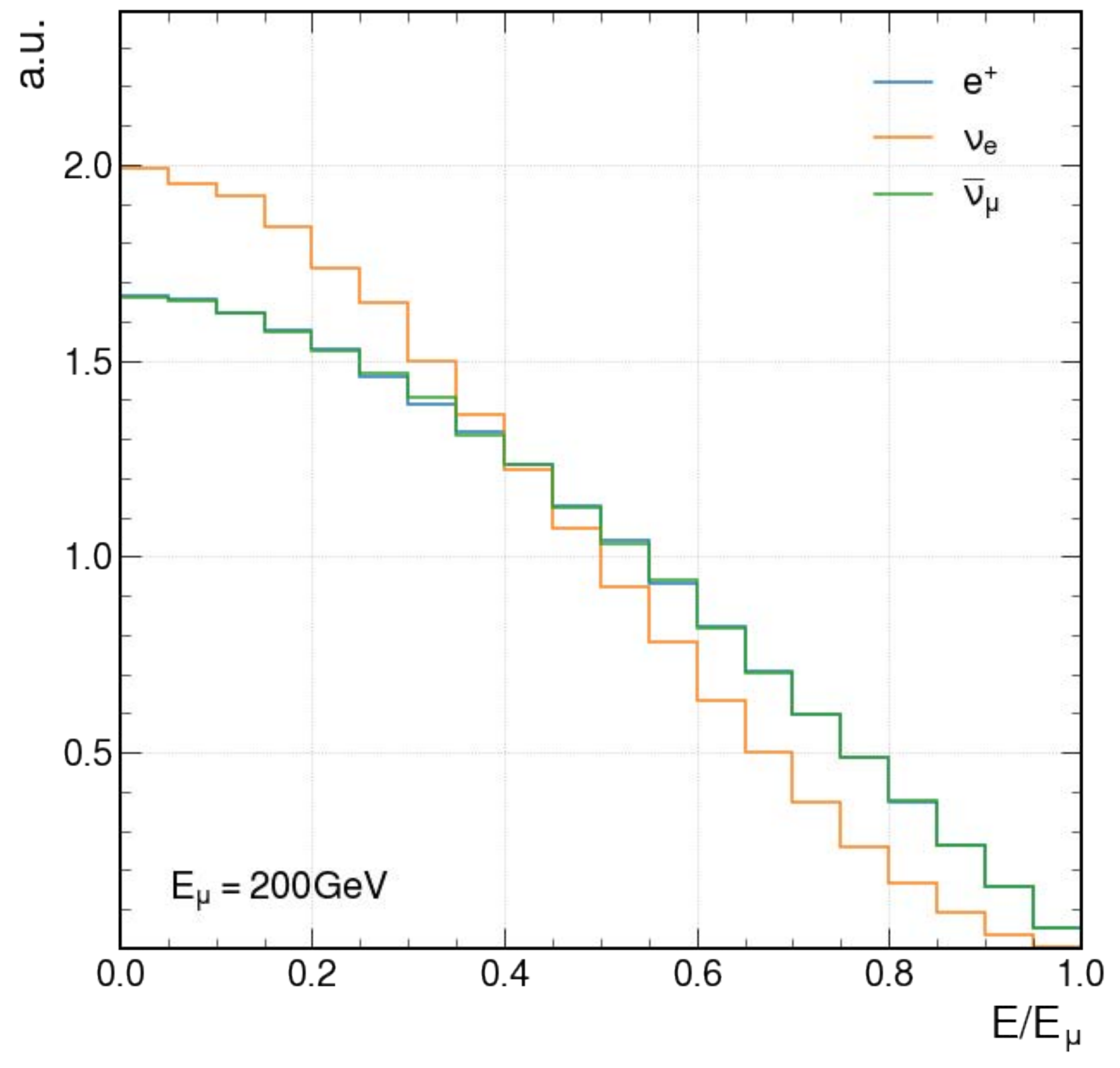}
    \caption{Energy fraction distributions of products emitted from 1000 GeV and 200 GeV muon beams.}
    \label{fig:Eneu}
\end{figure}

The physics potential of novel collider schemes is elaborated in a step-by-step manner. We start with using a TeV scale $\mup\rightarrow  \elp\nue\nuam$ beam. Fig.~\ref{fig:Eneu} shows the energy distributions of muon decay products from a muon beam with energy at $1\TeV$ and 200\GeV. As the decay angle $\theta$ goes like $\theta\sim 10^{-4}/{\rm E(TeV)}$, the muon decay products will be more collimated with increasing beam energy~\cite{C24,C25}. With TeV scale $\mup\rightarrow \elp\nue\nuam$ and $\mum\rightarrow  \elm\nuae\num$ beams from two sides, there appears the collisions 
\begin{align}
&\nue \nuae \rightarrow \PZ \rightarrow \mup\mum. \label{eq:p0}
\end{align}

\begin{figure}
    \centering
    \includegraphics[width=0.6\textwidth]{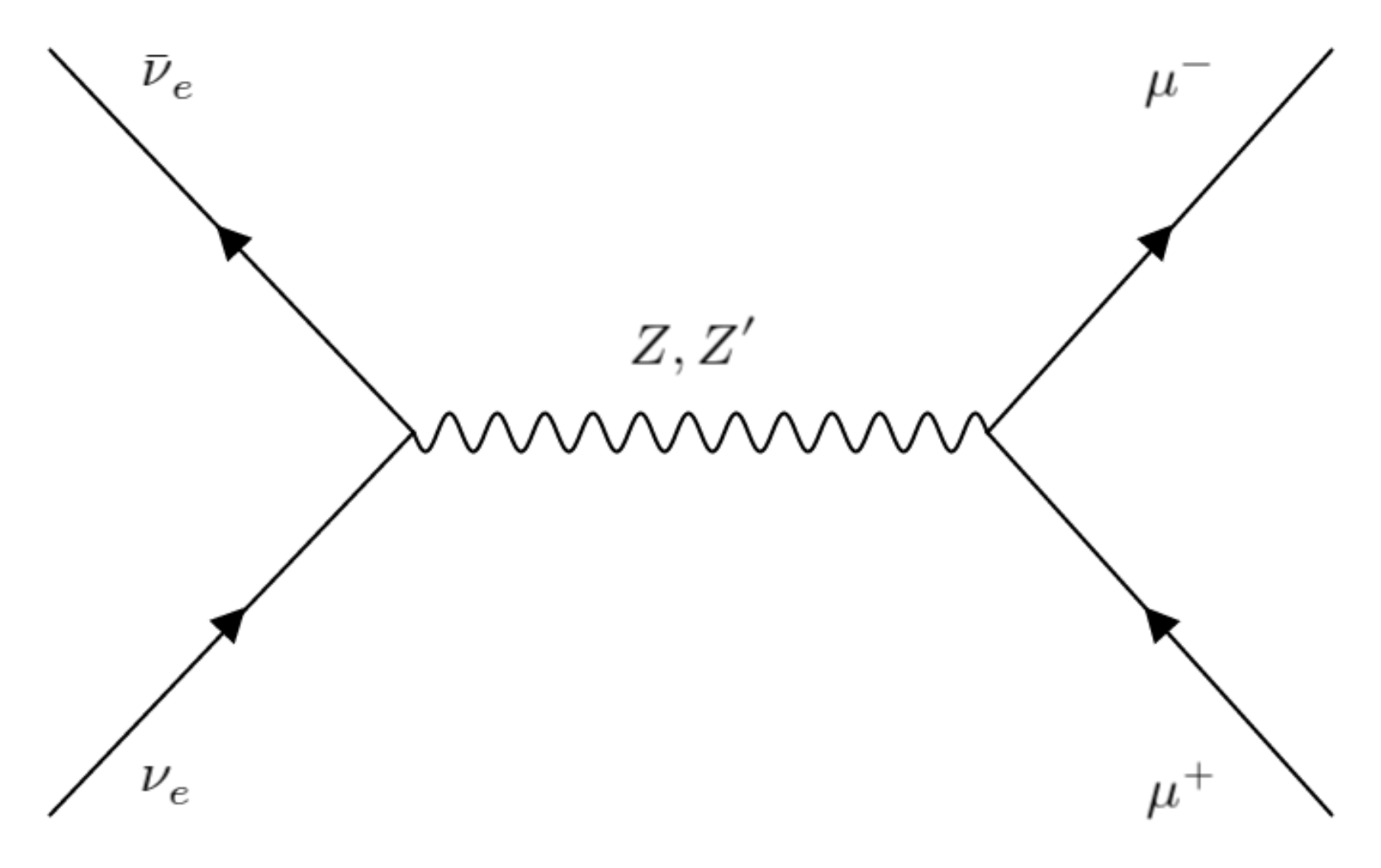}
    \caption{Example diagrams for neutrino antineutrino annihilation into Z boson and SM-like Z' boson.}
    \label{fig:fdneu}
\end{figure}
Relevant Feynman diagrams for neutrino antineutrino annihilation into Z and SM-like Z' boson are shown as in Fig.~\ref{fig:fdneu}.

\begin{figure}
    \centering
    \includegraphics[width=0.66\textwidth]{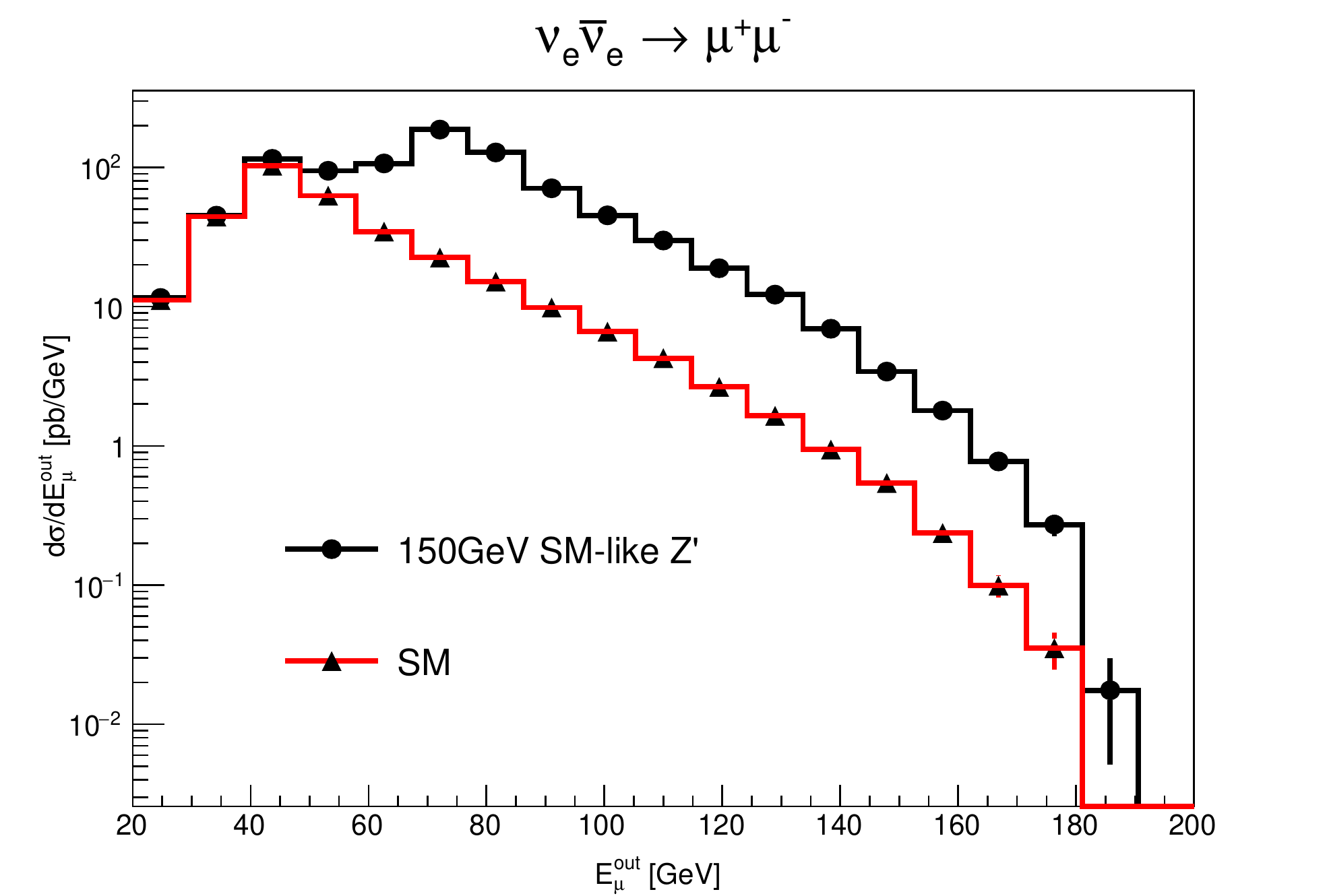}
    \caption{Outgoing muon energy distributions for neutrino antineutrino annihilation into Z and SM-like Z' bosons, with Z' mass set as $150\GeV$ and narrow width.}
    \label{fig:Emu}
\end{figure}

To simulate these processes, we implemented the neutrino energy fraction function~\cite{C24} from $200\GeV$ muon decay in \MGMCatNLO~\cite{C26}. The cross section reads 320 pb after requiring the final state muon to satisfy $\pt>20$\,GeV and $|\eta|<3.0$. As for the same process but with hadronic Z decay, the cross section will be around 5200 pb. Such large cross section can compensate for the luminosity limitation. For example, with a tiny integrated luminosity of about $10^{-5}$\fbinv, one can already expect to observe direct neutrino collisions through process.~\ref{eq:p0} and further can probe the $\PZ\nu\bar{\nu}$ couplings ~\cite{C27}, or search for possible $\nu\bar{\nu}$ resonance. Figure.~\ref{fig:Emu} shows the outgoing muon energy distributions for neutrino antineutrino annihilation into Z and SM-like Z' bosons, with Z' mass set as $150\GeV$ and narrow width.

On the other hand, we can also consider neutrino neutrino collision, which is specifically sensitive to the Weinberg operator and Majorana neutrino mass. With $1\TeV$ $\mup\rightarrow  \elp\nue\nuam$ beams from two sides, some of the main physics processes can be shown as below (where we only consider $\nue$ for simplicity):
\begin{align}
&\nue \nue \rightarrow \PH\PH \label{eq:p1}\\
&\nue \nue \rightarrow \PZ\PZ\,, \PZ\PH \label{eq:p2}\\
&\nue \nue \rightarrow \nue \nue \PH, \label{eq:p3}\\
&\nue \nue \rightarrow \nue \nue \PZ\PZ\,, \nue\nue\PW\PW, \label{eq:p4} \\
&\nue \nue \rightarrow \nue \nue \PZ\PH\,, \nue \nue \PH\PH, \label{eq:p5} \\
&\nue \nue \rightarrow e^-e^- \PW^+ \PW^+, \label{eq:p6}
\end{align}
The first two are generated using the so called Phenomenological Type I Seesaw model~\cite{C28}, with HMN mass scanned from several to hundred TeV, and the HMN mixing element set as $V_{eN}=0.01$ (where N stands for neutrino) by default. The latter three arise from so called vector boson fusion processes in the SM. The cross sections read:  11 fb for process.~\ref{eq:p1} and negligibly small for process.~\ref{eq:p2}, with HMN mass $\mn=20$\,TeV;  133 fb for process.~\ref{eq:p3}; 14 fb for process.~\ref{eq:p4};  0.17fb for process.~\ref{eq:p5}; 27 fb for process.~\ref{eq:p6} after requiring the final state electron to satisfy $\pt>10$\,GeV and $|\eta|<5.0$. 

Based on a simulation study~\cite{C24}, we expect for 20 TeV HMN, the 95\% C.L. limit on $V_{eN}$ can be set around 0.01, which surpasses current best results from the CMS experiments by two orders of magnitude~\cite{C29}.

With a TeV scale $\mup\rightarrow  \elp\nue\nuam$ beam, while the collision beams from the other side are $\elm$, $\elp$ and $\mum$, respectively, some of the main physics processes can be shown as below:
\begin{align}
&\elp \elm \rightarrow \PZ^{0(*)},\,\,\, \nue \elm \rightarrow \nue \elm,\,\,\, \nuam \elm \rightarrow \nuam \elm, \\
&\nue \elp \rightarrow \PW^{+(*)},\,\,\, \nuam \elp \rightarrow \nuam \elp,\,\,\, \nuam \elp \rightarrow \nuae \mup, \\
&\nuam \mum \rightarrow \PW^{-(*)},\,\,\, \nue \mum \rightarrow \nue \mum,\,\,\, \nue \mum \rightarrow \elm \num .
    \label{eq:mupdecay}
\end{align}

\begin{figure}
    \centering
    \includegraphics[width=0.46\textwidth]{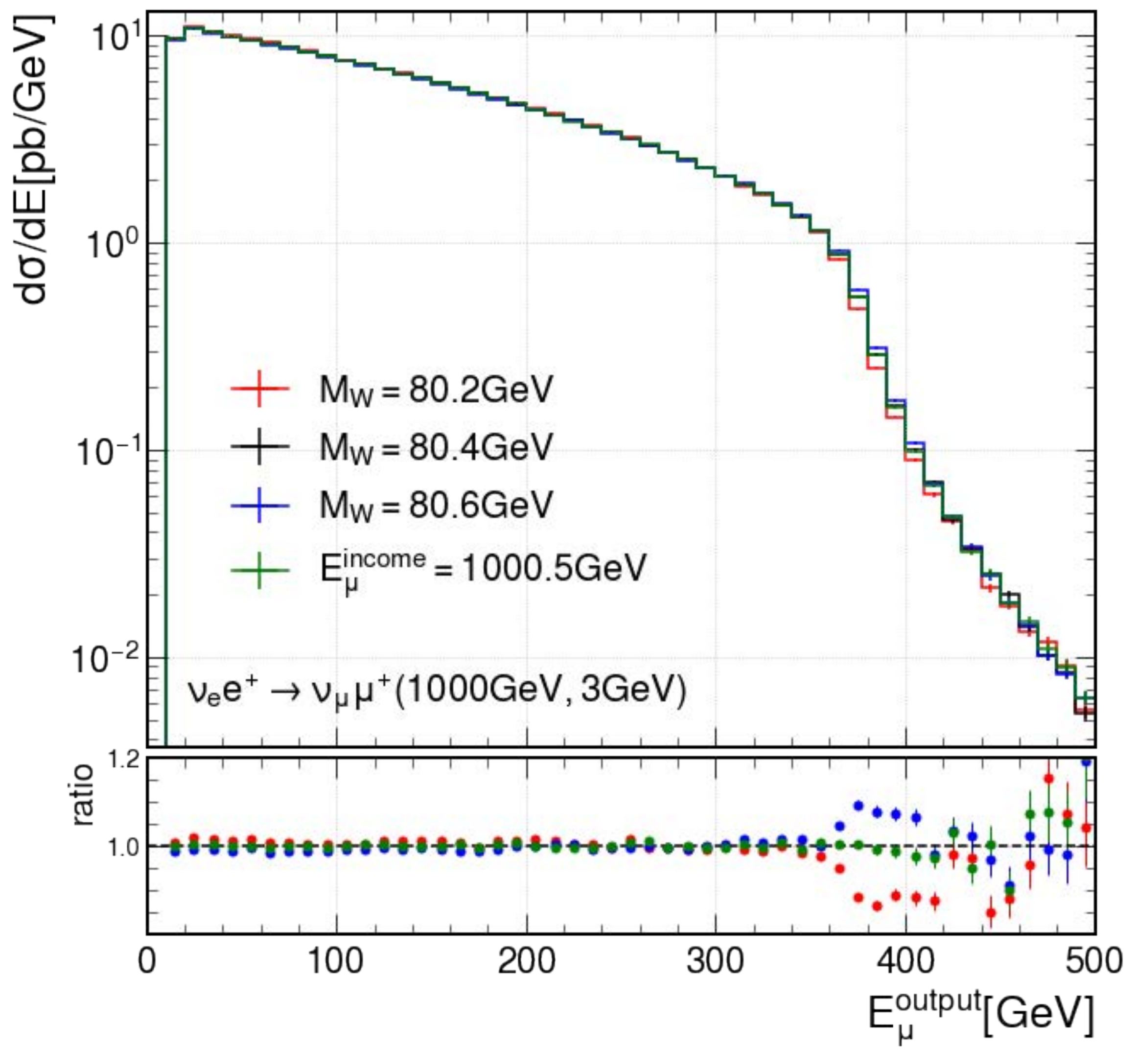}
    \includegraphics[width=0.46\textwidth]{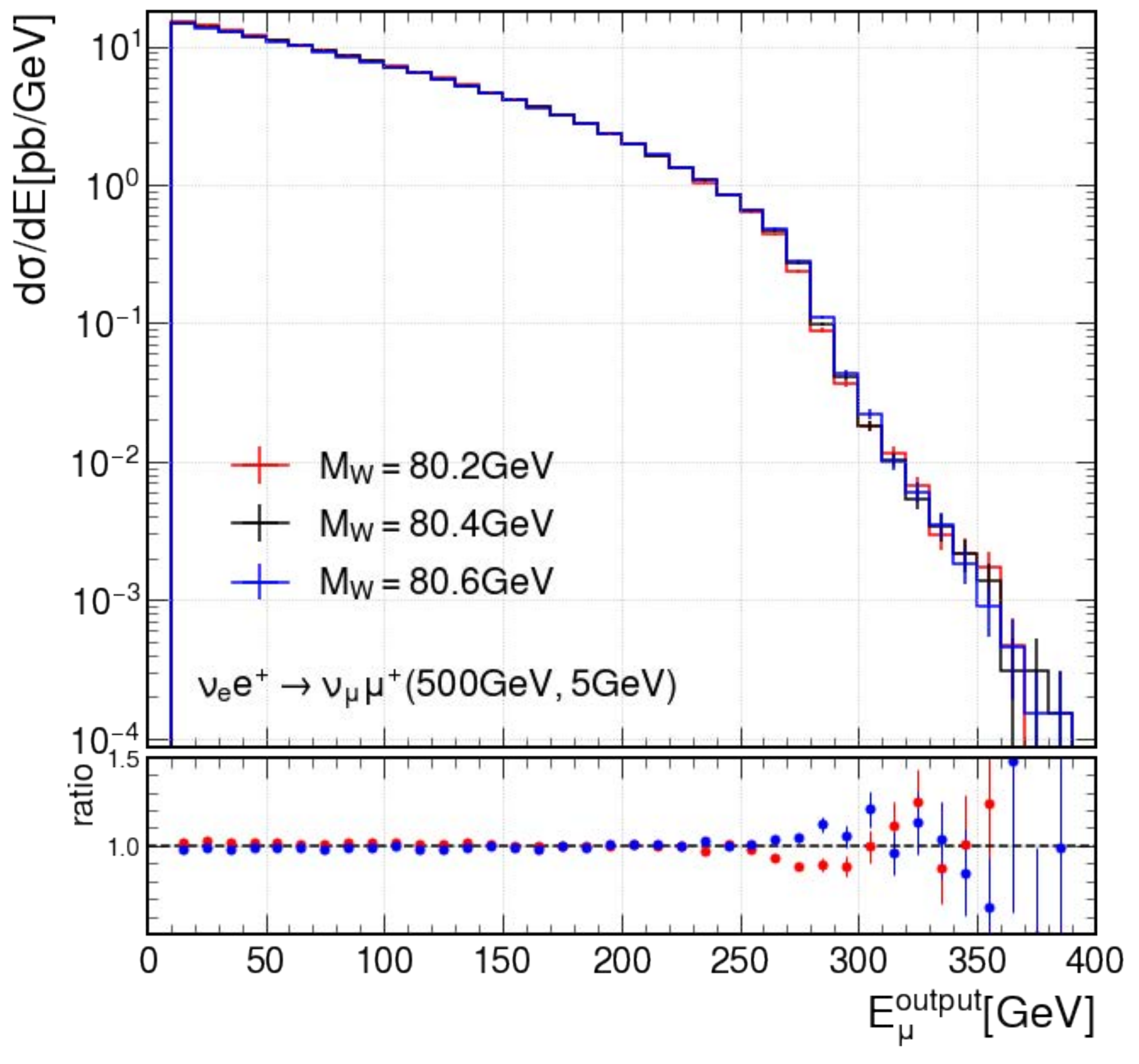}
    \caption{Distributions of the outgoing muon energy from $\nue \elp \rightarrow \PW^{+(*)} \rightarrow \num\mup$ at two collision scenarios: a neutrino beam arising from a 1000 (500) GeV muon beam, and a 3 (5) GeV positron beam. Clearly visible differences are seen between the $\mathrm{M_W}=80.2$, $80.4$ and $80.6$\,GeV cases. The figure above also shows the energy comparison of output muon when the income muon energy of 1000 GeV varies by 0.5 GeV at $\mathrm{M_W}=80.4\mathrm{GeV}$. Ratios are defined as distributions in other colors divided by the distribution in black (${\rm E}_{\mu}^{\mathrm{income}}=$1000 (500) GeV and $\mathrm{M_W}=80.4\mathrm{GeV}$). Error bars include only the statistical errors.}
    \label{fig:emu}
\end{figure}

We are especially interested in $\nue \elp \rightarrow \PW^{+(*)}$, which has a cross section that depends on $\mathrm{M_W}$. Based on a simulation study done in Ref.~\cite{C24} and as shown in Fig.~\ref{fig:emu}, we expect a 10 MeV accuracy on $\mathrm{M_W}$ can be achieved with an integrated luminosity of only 0.1\fbinv, which is already comparable with the CDF measurement result~\cite{C18}. Other potential physics results from such a neutrino lepton collider include a search for leptophilic gauge bosons, and studies of neutrino scattering processes to probe the $\PZ\nu\nu$ couplings.

\begin{figure}[htbp]
  \begin{center}
  \includegraphics[width=0.6\textwidth]{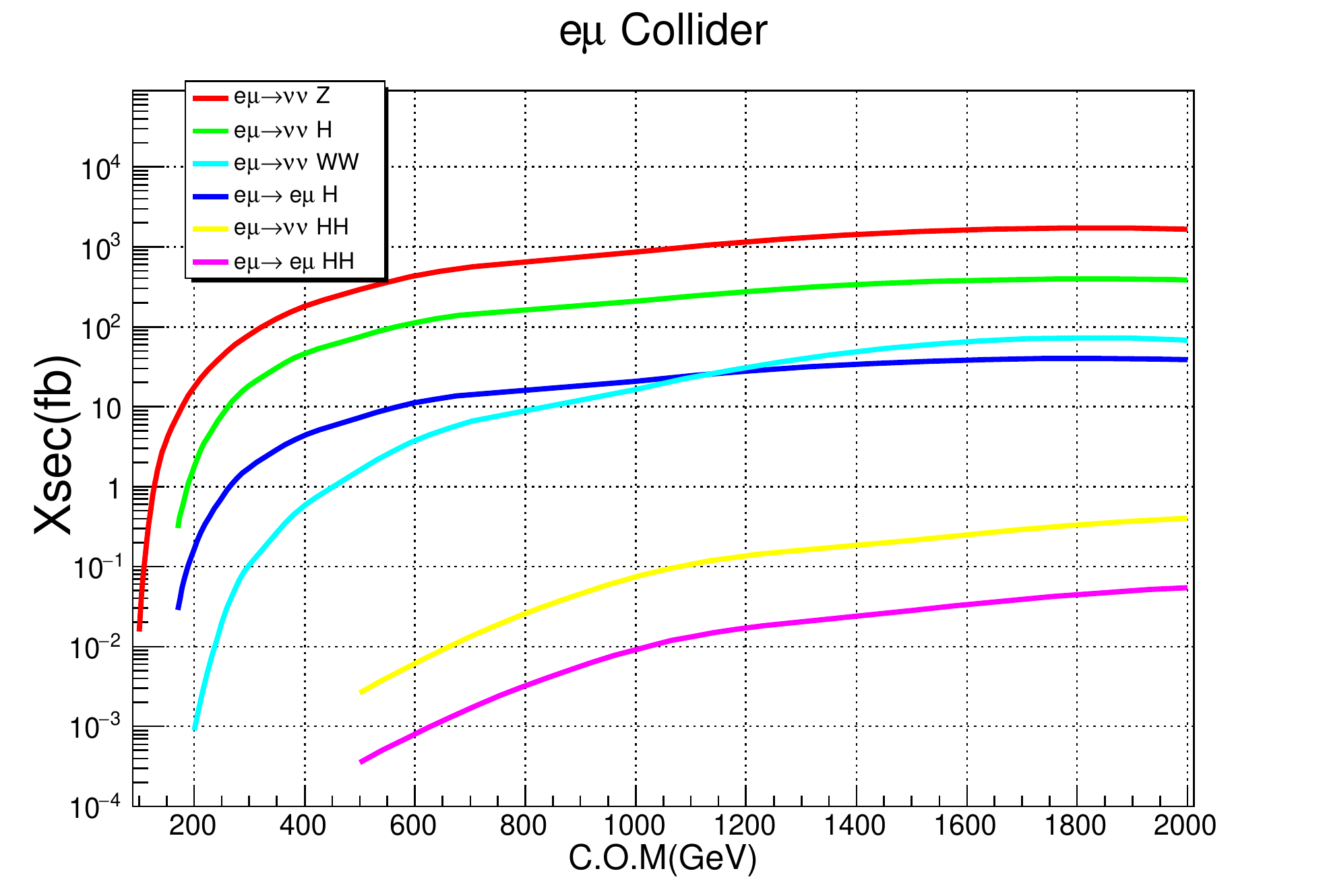}
    \caption{Cross section dependence on center-of-mass energy of six dominant physics processes at an electron-muon collider.}
    \label{fig:emucollider}
  \end{center}
\end{figure}

In addition to neutrino-neutrino and neutrino-lepton collisions, one can  extend to electron muon and muon muon collisions~\cite{C24}. The collision of an electron and muon beam leads to less physics background compared with either an electron-electron or a muon-muon collider, since electron-muon interactions proceed mostly through higher order vector boson fusion and vector boson scattering processes, as shown in Fig.~\ref{fig:emucollider}. The asymmetric collision profile results in collision products that are boosted towards the electron beam side, which can be exploited to reduce BIB from the muon beam to a large extent. 

With these discussions, one can imagine a lepton collider complex, starting from colliding order 10 GeV electron and muon beams for the first time in history and to probe charged lepton flavor violation, then to be upgraded to a collider with 50-100 GeV electron and 1-3 TeV muon beams to measure Higgs properties and search for new physics, and finally to be transformed to a TeV scale muon muon collider. The cost should vary from order 100 millions to a few billion dollars corresponding to different stages, which make the funding situation more practical (see also report on electron-muon collider from~\cite{C30}).  

All these novel types of collisions will give us rich phenomena unexplored before and a rich variety of physics goals could be achieved. For neutrino-neutrino or neutrino-lepton collisions,
only a tiny or small integrated luminosity is needed to yield wonderful physics results. 
We envision them as ``Richness out of smallness”, in partial answer to great P. W. Anderson’s ``More is different”. 

\section{Technical considerations}
\label{tc}

\begin{figure}
    \centering
    \includegraphics[width=.7\columnwidth]{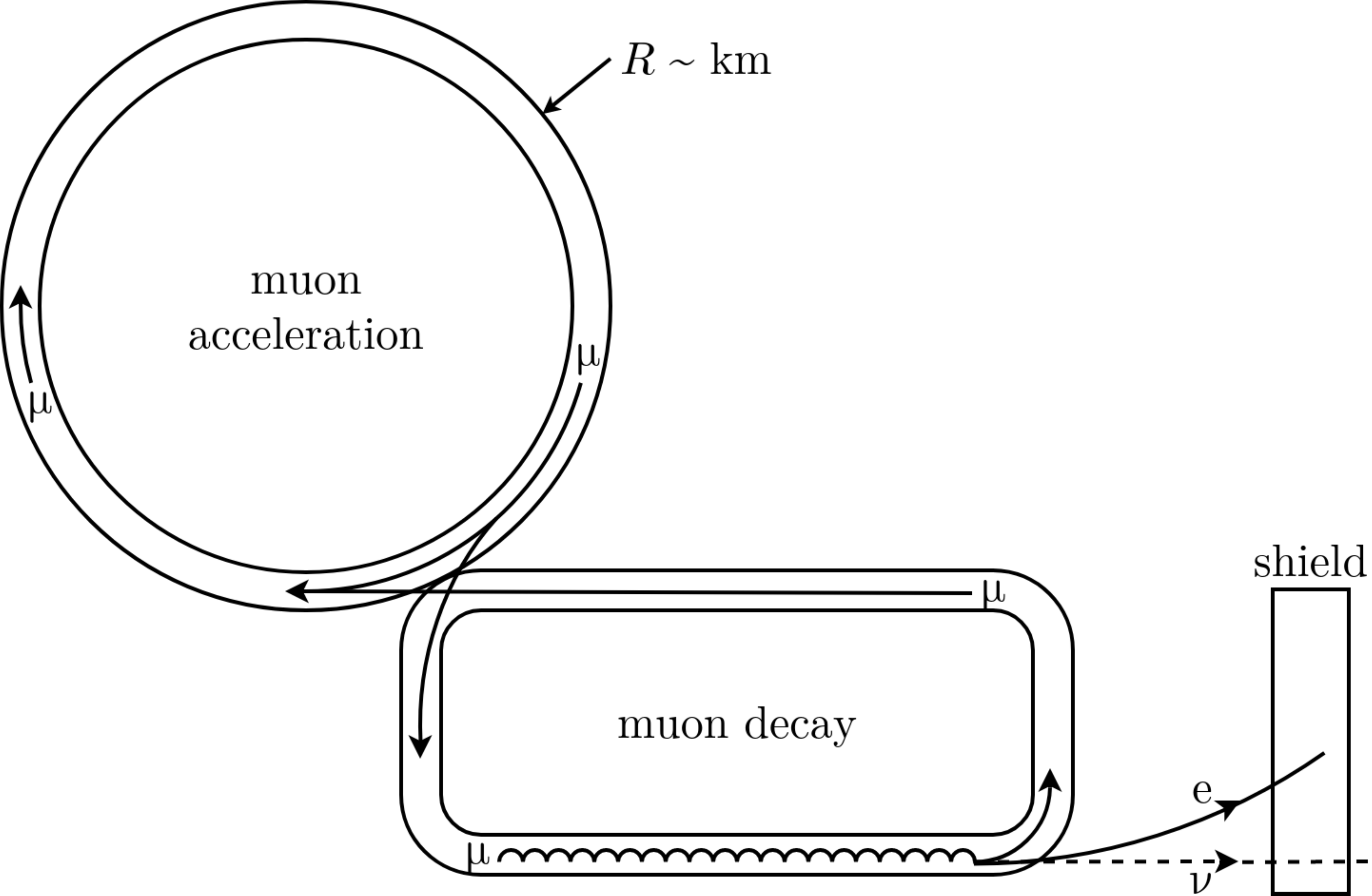}
    \caption{An illustration of the proposed neutrino beam and collider. Muons are accelerated in the circular section in the upper left, and then extracted into the rectangular section in the lower right. In one of the long edges of the rectangle, the neutrinos emitted from the muon decays are formed into a collimated beam. A small modulation of the muon decay angle through vertical bending, symbolized by the squiggly line, is used to focus the neutrino beam.}
    \label{fig:design}
\end{figure}

A neutrino collision complex from high-energy muon beams was discussed in Refs.~\cite{C24,C25}. The illustration of the proposed neutrino beam and collider is shown in Fig.~\ref{fig:design}. The muon beam is accelerated in the circular section in the upper left and then extracted into the rectangular section in the lower right. During each cycle, the muon beam will be squeezed due to Lorentz contraction and then pass through an arc ($L_c$) and linear sections ($L_l$), and emit bunches of collimated neutrinos finally. The electrons from the muon decays can either be shielded or used for energy calibration through collision with positrons from the other side. Using the two rings instead of one ring here allows for more flexibility to accommodate crowded bunches with different time or space gaps.

The instantaneous luminosity of a neutrino collider would be limited by two main factors: 1) the intensity of the neutrino beam compared with the incoming muon beam is suppressed by roughly $L_l/L_c\sim 0.1$, i.e., the fraction of the collider ring circumference occupied by the production straight section~\cite{C25}, 2) the neutrino beam spread, which may still be kept at 10 to 100 microns at the interaction point, by applying a small modulation on muon decay angle through vertical bending to achieve more focused neutrino beam~\cite{C35}.

We provide more details using the formula for the instantaneous luminosity,
\begin{align}
  {\cal L} = {N_{\rm beam 1} N_{\rm beam 2} \over 4 \pi \sigma_x \sigma_y} f_{\rm rep},
\end{align}
where $f_{\rm rep}$ is the rate of collisions and is typically 100 kHz (40 MHz) for lepton colliders (hadron colliders), and $N_{\rm beam 1,2}$ are the number of particles in each bunch which can be taken as $\sim 10^{11}\text{--}10^{12}$~\cite{ILCtdr,FCCee,CEPC}, $\sigma_x$ and $\sigma_y$ are the beam sizes. Take the LHC as an example~\cite{lhclumi}, with $f_{\rm rep}=40$\,MHz, $\sigma_{x,y}=16$ microns, and $N_{\rm beam 1,2}=10^{11}$, one can get $ {\cal L}=10^{34}$ cm$^{-2}$s$^{-1}$. As for TeV muon colliders~\cite{Delahaye:2019omf,Bossi:2020yne}, with $f_{\rm rep}=100$\,KHz, $\sigma_{x,y}\lesssim 10$ microns, and $N_{\rm beam 1,2}=10^{12}$, then $ {\cal L}=10^{33}\text{--}10^{34}$ cm$^{-2}$s$^{-1}$. 
As for the neutrino neutrino collisions discussed above, there are further suppression factors from linear over arc ratio ($L^2_l/L^2_c\sim 1/100$) with the exact value depending on the realistic design as shown in Fig.~\ref{fig:design}, and the neutrino beam spread which can be around $r_s\sim 1000$ microns for $L_l\sim$ 10 to 100 meters. Taking all these into account, a realistic instantaneous luminosity for neutrino neutrino collisions can reach around ${\cal L}=10^{28}$ cm$^{-2}$s$^{-1}$ level. Although it is a small number, however, to reach the discovery threshold of neutrino antineutrino annihilation process $\nue \nuae \rightarrow \PZ$, a tiny integrated luminosity of about $10^{-5}$\fbinv~is needed, i.e., several days of data taking. 

\begin{figure}
    \centering
    \includegraphics[width=.7\columnwidth]{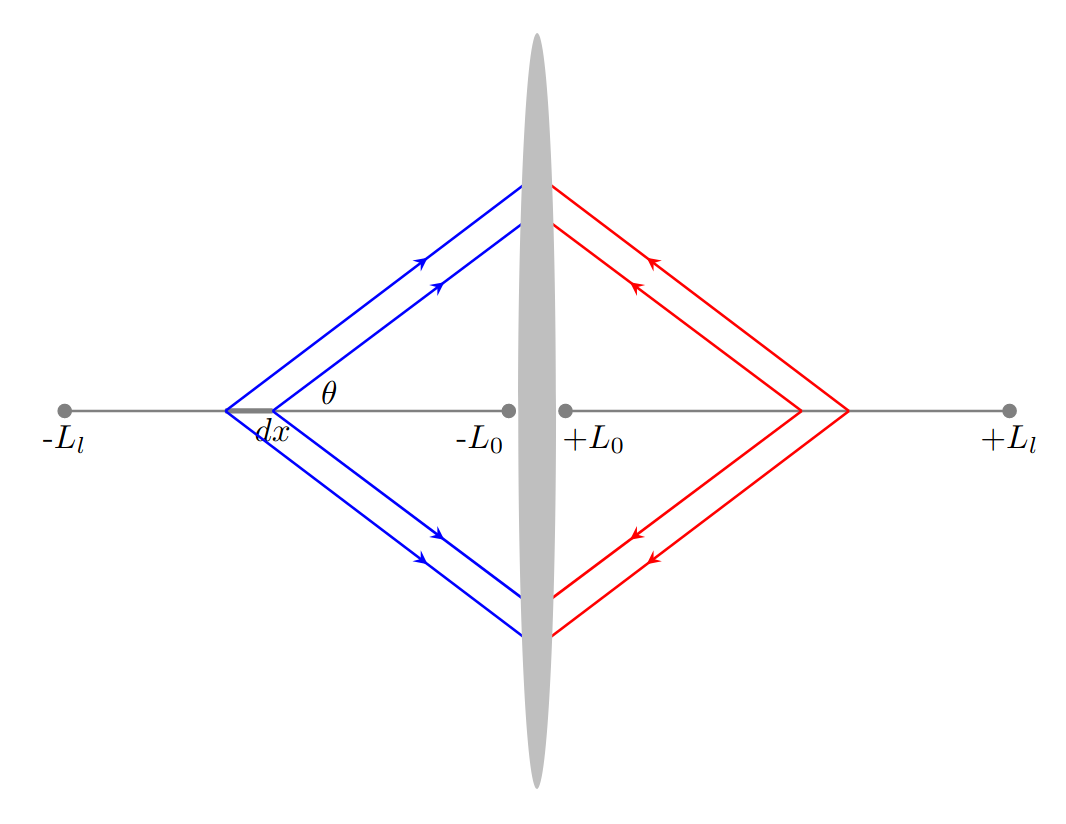}
    \caption{The neutrino neutrino collision profile,  where muon beams flying the linear path radiate neutrinos approximately along a cone with polar angle as $\theta$.}
    \label{fig:lumi}
\end{figure}

To be more practical, we consider the neutrino neutrino collision profile as shown in Fig.~\ref{fig:lumi},  where muon beams flying the linear path (symbolized by $L_l$ to $L_0$, where $L_0$ is a cut-off parameter defined by the muon beam size
) radiate neutrinos approximately along a cone with polar angle as $\theta$. We then estimate the instantaneous luminosity for neutrino neutrino collisions as below:
\begin{align}
  {\cal L} &= \frac{L^2_l}{L^2_c} \int^{L_l}_{L_0}\frac{N_{\rm beam 1}  N_{\rm beam 2} f_{\rm rep}}{L^2_l\times (4\times 2\pi x\tan^2\theta)}\times dx \nonumber\\
  &= \frac{L^2_l}{L^2_c}\frac{N_{\rm beam 1} N_{\rm beam 2} f_{\rm rep}}{8\pi L^2_l\tan^2\theta}\times \ln(L_l/L_0),
\end{align}
with $L_l\tan\theta\sim r_s$, and there appears as an enhanced factor of $\ln(L_l/L_0)/2\sim 2-5$, and thus can further increase the instantaneous luminosity for neutrino neutrino collisions. Note $L_0$ is a cut-off parameter in above integration formula and defined by the muon beam size, which can be at the order of 1-10 cm and thus may relax the stringent requirement on beam cooling of nominal muon collider being pursued.

The lepton beam from the other collision side is of lower energy with a few GeV, and the quality can be improved by many high-current high-frequency techniques. Assuming the neutrinos emitted by TeV scale muon beams and electron energies around 5 GeV, the instantaneous luminosity can reach $10^{32}$ to $10^{33}\ {\rm{cm}}^{-2}{\rm{s}}^{-1}$ level. In the following study, we assume the integrated luminosity to be around 10 to 100 \fbinv~in 10 years.
The interesting physics process from neutrino antineutrino annihilation has a huge production cross-section and is expected to be observed with 200 GeV muon beam and a very tiny integrated luminosity of data. Therefore the requirement of the muon collider on beam quality can be relaxed to a large extent.

As for the electron muon and muon muon colliders~\cite{C24}, the physical infrastructure needed to generate electron and muon beams with energies below 100\,GeV would be either a small linear or circular accelerator. A muon beam at TeV scale can be achieved in a kilometer-sized storage ring. 
Specifically to our interest, 200 GeV and 1 TeV muon beams are crucial for two important physics processes, i.e., neutrino-antineutrino annihilation into Z boson, and neutrino-neutrino scattering through lepton flavor violation (LFV) process into two Higgs, respectively.

A set-up that can generate both electron and positron beams, or both muon and anti-muon beams, or polarized beams, should be considered to further enrich the physics outcome. 
Such a lepton collider complex can start from colliding order 10 GeV electron and muon beams for the first time in history and probe charged LFV (CLFV), then be upgraded to a collider with 50--100 GeV electron and 1-3 TeV muon beams to measure Higgs properties and probe BSM physics including CLFV, and finally to be transformed to a TeV scale muon-muon collider. Note such a lepton collider complex can also serve as a muon source at various energy scales for general scientific research.
  
\section{Status and Prospects}
\label{statusprospects}

Muon beams are usually achieved in proton or positron interactions on target, although there have been progress made recently~\cite{C32,C33,C34}, which are still facing challenges to overcome on  beam cooling or intensity. As a countermeasure, we can also consider those important physics goals with low luminosity requirements.

Several related neutrino scattering experiments have been proposed in the last few decades, including NuTeV~\cite{C37}, NuMAX~\cite{C38}, NuSOnG~\cite{C39}, and nuSTORM~\cite{C40}. Their motivations include, e.g., making precision neutrino interaction cross section measurements, or searching for neutrino related non-SM physics. However, a head-on neutrino lepton collider at the 100 GeV scale is proposed here for the first time, with rich physics potential.


We are planning to work closely with the international Muon Collider Collaboration~\cite{C31} and closely follow domestic and international development on muon physics.

The underlying Deliberation Document~\cite{C36} of the 2020 Update of the European Strategy for Particle Physics stipulates that, in addition to high field magnets, the accelerator R$\&$D roadmap should contain: ``…an international design study for a muon collider, as it represents a unique opportunity to achieve a multi-TeV energy domain beyond the reach of $e^+e^-$ colliders, and potentially within a more compact circular tunnel than for a hadron collider. The biggest challenge remains to produce an intense beam of cooled muons, but novel ideas are being explored''. The international Muon Collider Collaboration~\cite{C31} is now leading the R$\&$D study,  to establish whether the investment into a full Conceptual Design Report (“CDR”) and demonstrator for a muon collider is scientifically justified. The collaboration consists of tens of renowned institutes and universities over the world, and is aiming to finalize this study in 5-10 years. In particular, the study is focused on the high-energy frontier and considers options with a centre-of-mass energy of 3 TeV and of 10 TeV or more. Potential synergies with other projects are also being pursued.

Our program currently and for the next 2-3 years will be focusing on R$\&$D, especially physics feasibility study and simulation software. On beam, hardware and luminosity targets, we are following closely related colliders' designs (including CEPC and muon colliders). One of the key points of this program is on simulation and validation of muon decay secondary beams, especially on luminosity optimization over neutrino beams. Once fixing the muon source and acceleration options, we will try first to realise 100-200 GeV muon beam and secondary neutrino beams' collision. We anticipate to achieve bunches of important physics results with only a tiny data of $10^{-5}$\fbinv. Afterwards, we will upgrade to 1 TeV muon beam and neutrino collisions. 

We have already achieved a series of physics results with attention, on neutrino neutrino collisions, neutrino lepton collisions, electron muon and muon muon collisions~\cite{C24}. 
As for the physics analysis parts, one can use the Monte-Carlo generators such as MadGraph and Pythia to simulate the events at parton and hadron level, and cross-checked with the Whizard generator widely used by the CEPC experiment. Afterwards, fast simulations with Delphes or full simulations with GEANT4 and FLUKA will be studied in detail, which will also help detector designs for the novel type of colliders.

\section{Summary}
\label{summary}

In this document, we propose a staged blueprint on future High Energy Colliders: from neutrino neutrino collision,  neutrino lepton collision, to electron muon and muon muon collisions. 

Our physics target and development strategy are of novelty and may enjoy the first mover advantage. In particular, we propose for the first time the concept of neutrino neutrino collisions, which can serve as a pre-step of a high energy muon collider. To be more practical, one may start from neutrino collisions:  notice a tiny or small integrated luminosity is needed for neutrino neutrino or neutrino lepton collisions to achieve wonderful physics results as described above. Accordingly, the requirement on muon beams may not be so stringent.

We will work closely with international and domestic programs on muon source and acceleration, including the CSNS/EMus, HIAF and CiADs, etc~\cite{C36}. On the other hand, as the requirement on muon beam quality is relatively loose for neutrino collisions , we can also exploit the position driven muon source method, LEMMA~\cite{C33}. Besides, Chinese scientists have made many original and important contributions on several key technologies, such as proton target, slow muon and muonium generation, etc~\cite{C36}.

We anticipate our program can sustain for over 50 years. Due to the program's large scale, the most important tasks in the next few years will be finalizing a detailed conceptual white book.
We believe such a (preparatory) Laboratory on novel colliders could attract vast international interests and collaborations. 

\appendix

\begin{acknowledgments}
This work is supported in part by the National Natural Science Foundation of China under Grants No. 12150005, No. 12075004 and No. 12061141002, by MOST under grant No. 2018YFA0403900.
\end{acknowledgments}

\bibliographystyle{ieeetr}
\bibliography{h}
\end{document}